\documentclass[conference]{IEEEtran}
\IEEEoverridecommandlockouts
\usepackage{cite}
\usepackage{amsmath}
\usepackage{algorithmic}
\usepackage{graphicx}
\usepackage{textcomp}
\usepackage{xcolor}
\usepackage{adjustbox}
\usepackage{graphicx}
\usepackage{amssymb}
\usepackage{lineno}
\usepackage{mathtools}
\usepackage{hyperref}
\usepackage{mathrsfs}

\def\BibTeX{{\rm B\kern-.05em{\sc i\kern-.025em b}\kern-.08em
    T\kern-.1667em\lower.7ex\hbox{E}\kern-.125emX}}
\begin{document}

\title{Frequency and Time Domain Analysis of sRNA-based Treatment for Inflammatory Bowel Disease*\\{\footnotesize \textsuperscript{*}}
\thanks{BBSRC, Wellcome Trust, IDT, Eppendorf, BioLab, Snap gene and Oxford university departments of biochemistry, Engineering and chemistry}
}

\author{\IEEEauthorblockN{1\textsuperscript{st} Arman Karshenas}
\IEEEauthorblockA{\textit{Department of Engineering Science} \\
\textit{University of Oxford}\\
Oxford, UK \\
arman.karshenasnajafabadi@balliol.ox.ac.uk}
\and
\IEEEauthorblockN{2\textsuperscript{nd} Joseph Windo}
\IEEEauthorblockA{\textit{Department of Biochemistry} \\
\textit{University of Oxford}\\
Oxford, UK \\
joseph.windo@lmh.ox.ac.uk}
\and
\IEEEauthorblockN{3\textsuperscript{rd} Bhuvana Sudarshan}
\IEEEauthorblockA{\textit{Department of Biochemistry} \\
\textit{University of Oxford}\\
Oxford, UK \\
bhuvana.sudarshan@oriel.ox.ac.uk}
\and
\IEEEauthorblockN{4\textsuperscript{th} Jonathan Stocks}
\IEEEauthorblockA{\textit{Deptartment of Biochemistry} \\
\textit{University of Oxford}\\
Oxford, UK \\
jonathan.stocks@univ.ox.ac.uk}
\and
\IEEEauthorblockN{5\textsuperscript{th} adrian Kozhevnikov}
\IEEEauthorblockA{\textit{Department of Engineering Science} \\
\textit{University of Oxford}\\
Oxford, UK \\
Adrian.kozhevnikov@magd.ox.ac.uk}
\and
\IEEEauthorblockN{6\textsuperscript{th} Jhanna Kryukova}
\IEEEauthorblockA{\textit{Department of Biochemistry} \\
\textit{University of Oxford}\\
Oxford, UK \\
zhanna.kryukova@st-hughs.ox.ac.uk}
\and
\IEEEauthorblockN{7\textsuperscript{th} Eleanor Beard}
\IEEEauthorblockA{\textit{Department of Medicine} \\
\textit{University of Oxford}\\
Oxford, UK \\
eleanor.beard@st-annes.ox.ac.uk}
\and
\IEEEauthorblockN{8\textsuperscript{th} Laurel Constanti Crosby}
\IEEEauthorblockA{\textit{Department of Biology} \\
\textit{University of Oxford}\\
Oxford, UK \\
laurel.constanticrosby@stcatz.ox.ac.uk}
\and
\IEEEauthorblockN{9\textsuperscript{th} Max Taylor}
\IEEEauthorblockA{\textit{Department of Biochemistry} \\
\textit{University of Oxford}\\
Oxford, UK \\
max.taylor@st-hildas.ox.ac.uk}
}

\maketitle

\begin{abstract}
The validity of a complex reaction pathway proposed to treat Inflammatory Bowel Disease (IBD) was verified by a comprehensive time and frequency domain analysis. The model was taken to the frequency domain to study the effect and the significance of the negative feedback loop introduced by the reaction pathways. It could be shown that such proposed probiotics have very interesting potentials that could be used extensively in near future. 
\end{abstract}


\section{Introduction}
\begin{figure}[h]
    \centering
    \includegraphics[width = 8.5 cm , height = 4.5  cm]{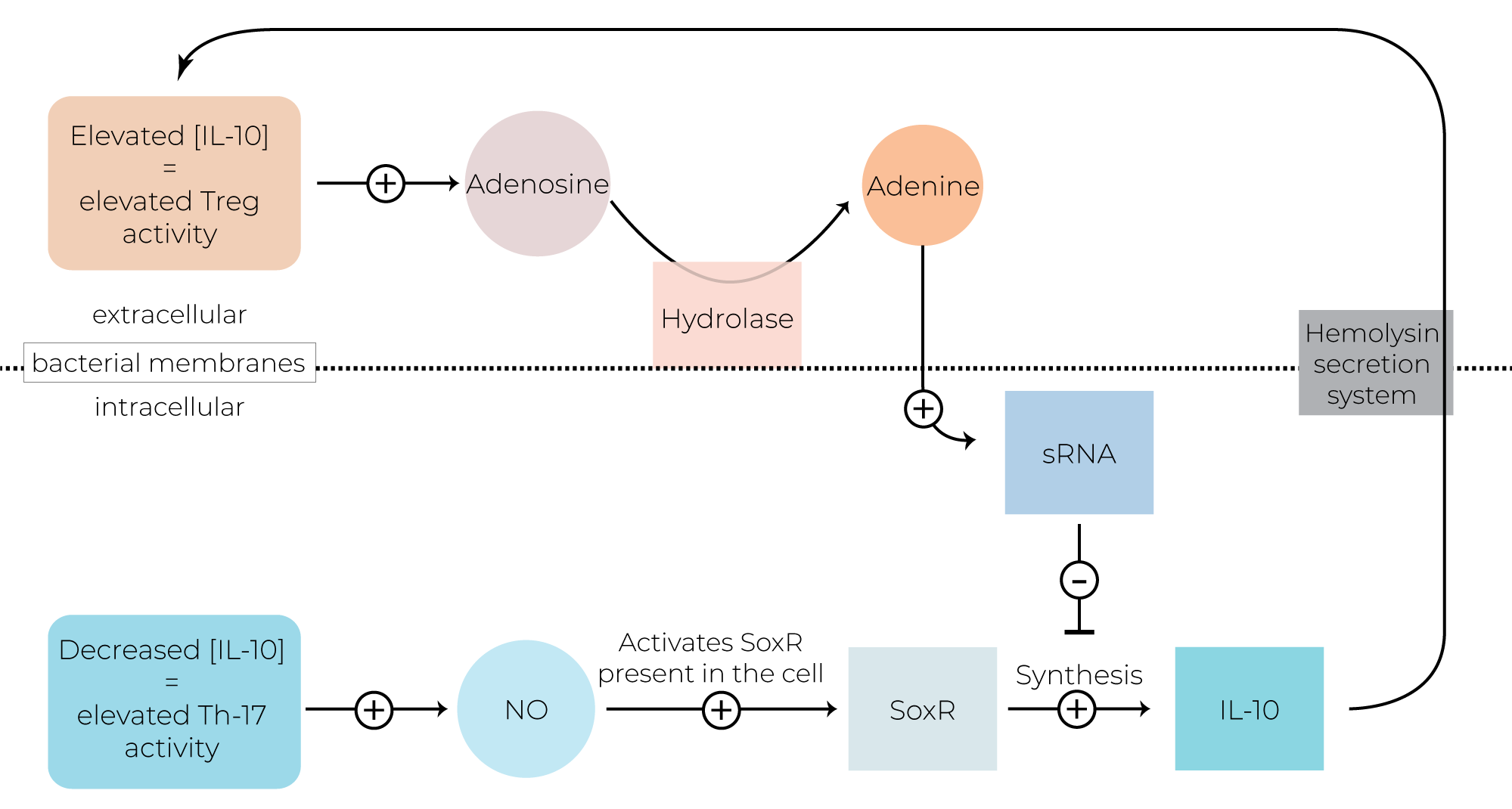}
    \caption{Simplified schematic diagram of the negative and positive pathways}
    \label{partsDNA}
\end{figure}

Inflammatory Bowel Disease (IBD) is characterised by chronic inflammation of the intestine. The condition is associated with an imbalance in immune cell populations, notably Th17 and Treg. Existing immunosuppressive therapies, when successful, often elicit systemic side effects and require frequent re-administration. Therefore, the proposed solution is a probiotic strain that restores the Th17/Treg cell balance via secretion of IL-10 in response to Nitric oxide in the intestinal lumen. Overshoot is prevented by an adenine riboswitch-sRNA construct which responds to extracellular adenosine, an indicator of the Treg cell population. The two independent reaction pathways - responding separately to IL-10 deficiency and abundance - and are summarised in Figure~\ref{partsDNA}.  

To avoid the complexity of multivariate control systems, the pathways have been modelled independently. Physiologically, an elevated serum concentration of NO can be associated with autoimmunity, whereas an elevated serum adenosine concentration can signify immunoinsufficiency. The NO pathway will be referred to as the `negative pathway', representing a negative concentration difference in IL-10. Likewise, the adenosine pathway has been termed the `positive pathway'. Integration of separate stimuli in a dual feedback loop enables a more dynamic, robust response to the immune state of the body.  Thus the engineered probiotic bacteria would be capable of maintaining an equilibrium between the Th17 and Treg cell populations by virtue of the negative feedback system.
\subsection{Negative Pathway}
The negative pathway can be modelled in a canonical manner: nitric oxide triggers oxidation of the SoxR iron-sulphur cluster, allowing binding to the promoter of IL-10 in the circuit - pSoxS. This results in transcriptional activation of the system within approximately $0.01 \; $seconds ~\cite{Introtosysbio}. Hence, the system has been modelled by a simple Hill activation function for IL-10 gene transcription. The SoxR/pSoxS system has been studied extensively~\cite{soxr} but previous modelling has been limited to that of the SoxR/pSoxS binding interaction by the 2009 Stanford iGEM team \cite{stanford}.

\subsection{Positive Pathway}
The positive pathway mechanism facilitates translational inhibition of IL-10 to prevent overshoot. The anti-inflammatory molecule, adenosine, is produced by Treg cells in response to inflammation, with elevated levels indicating Treg overactivity and an immunoinsufficient state. Adenosine is subsequently broken down into ribose and adenine, via an extracytoplasmic, nucleoside hydrolase, by the engineered cell. Following uptake into the cell, adenine binds to a riboswitch, resulting in an increased rate of RNA transcription. A self-splicing ribozyme separates the riboswitch from the remaining transcript - an sRNA molecule which is complementary to the 5' region of the IL-10 mRNA. 

Consequently, translation of IL-10 mRNA is prevented by the binding of sRNA. It should be noted that NO is an essential input to the positive pathway and hence, combined model for both reaction pathways could be found below in Figure`\ref{schempaths}. 
\begin{figure}[h]
    \centering
    \includegraphics[width = 8.5cm ]{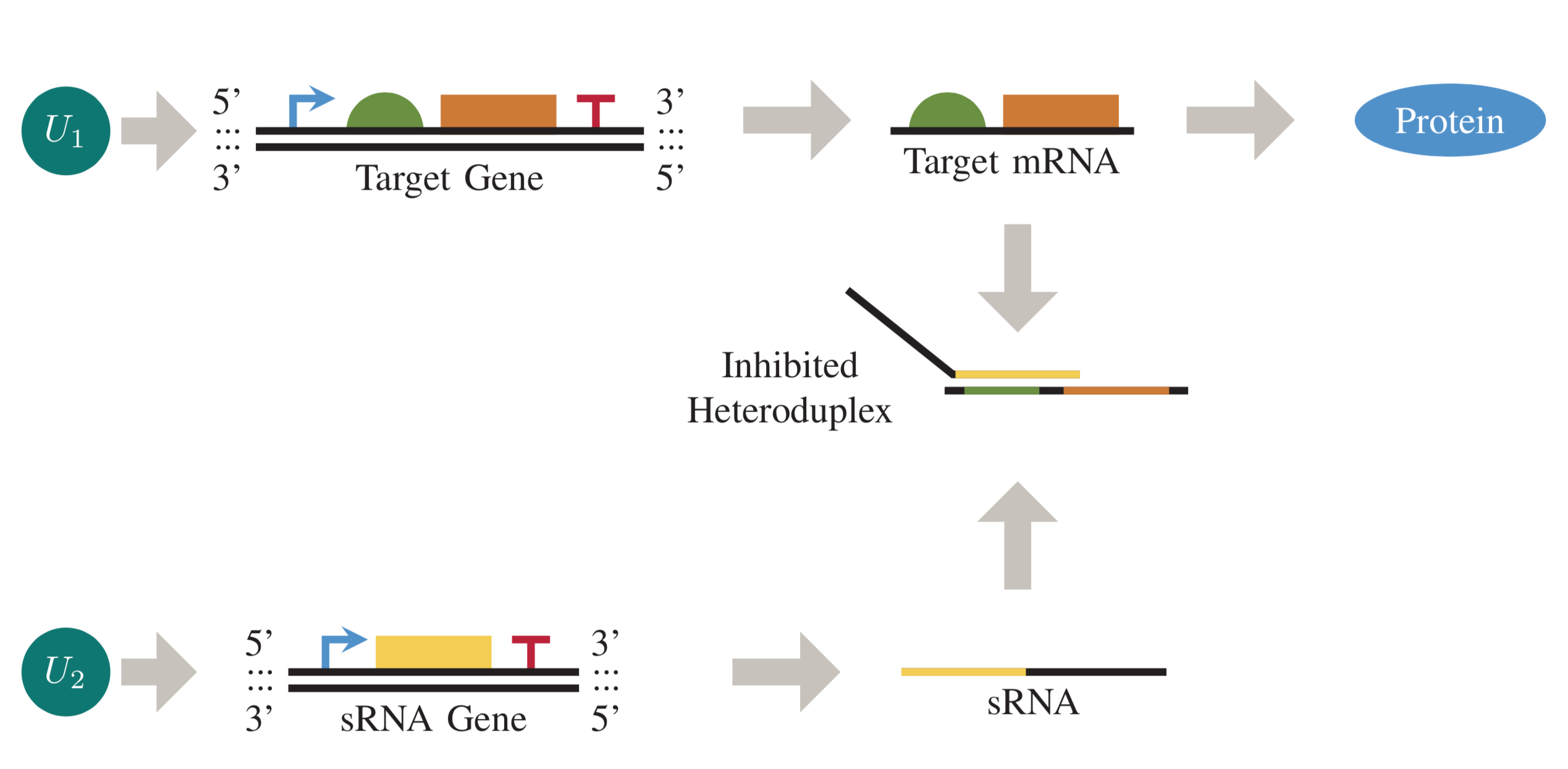}
    \caption{Schematic representation of the pathways \cite{oxfordeng}}
    \label{schempaths}
\end{figure}
\section{Dynamic Model}
The dynamic model of both of the pathways were analysed using the Simbiology toolbox in MATLAB as well as ODE15s Solver and some of the modelling and parameters are based on the paper "Frequency domain analysis of small non-coding RNAs"~\cite{oxfordeng}. The dynamic time domain analysis of the system was conducted using the combined model in Figure~\ref{schempaths} with different initial conditions corresponding to negative and positive pathways separately. It is worth mentioning that data fitting was used to determine some of the parameters. A list of the parameters and data is available at the end of paper. The differential equations describing the systems are as follows: 
\newline
\begin{equation}
\label{eq1}
    \frac{d[\text{sRNA}]}{dt} = \Gamma_1 - \alpha_1[\text{sRNA}] - k_1[\text{IL-10 mRNA}][\text{sRNA}],
\end{equation}

\begin{equation}
\label{mRNA}
    \frac{d[\text{IL-10 mRNA}]}{dt} = \Gamma_2 - \alpha_2[\text{IL-10 mRNA}] - k_1[\text{IL-10 mRNA}][\text{sRNA}],
\end{equation}
\begin{equation}
\label{eq3}
    \frac{d[\text{IL-10 - i}]}{dt} = k_2[\text{IL-10 mRNA}] - \alpha_3[\text{IL-10 - i}] - k_3[\text{IL-10 - i}],
\end{equation}
\begin{equation}
\label{eq4}
    \frac{d[\text{IL-10 - s}]}{dt} = k_3[\text{IL-10 - i}] - \alpha_4[\text{IL-10 - s}],
\end{equation}
\newline
where $\Gamma_i$ are the Hill functions describing the rate of transcription with inputs of Adenosine and NO for $\Gamma_1$ and $\Gamma_2$ respectively as represented by $u_i$ in (\ref{Hillfunction}).
\begin{equation}
    \Gamma_i = \frac{\beta [u_i]^n}{K^n+[u_i]^n},
    \label{Hillfunction}
\end{equation}
In which $\beta$ is the maximal transcription factor, $K$ the dissociation coefficient and $n$ the Hill coefficient. $k_1$ is the rate of sRNA binding to IL-10 mRNA which heavily depends on the length of the sRNA, making it easy to exploit this relationship for better control over the fate of the system. $k_2$ is the translation rate of IL-10 mRNA and $k_3$ is the secretion and diffusion rate of IL-10~\cite{Secretion}. $\alpha_i$ represents the dilution + degradation rate. 
\newline
It is important to notice that Hill functions introduce non-linearity into the system as well as sRNA binding. Therefore, Hill functions have been linearised for transfer function derivation in the frequency domain, and are noted by $\gamma^*$ as follows:
\begin{equation}
\label{linearhill}
   \gamma^*_i = \beta_i \frac{n_iK_i^{n_i}[u_i]^{n_i-1}}{(K_i^{n_i}+u_i^{n_i})^2},
\end{equation}
All the assumptions made for (\ref{eq1}) to (\ref{eq4}) are listed below:
\begin{itemize}
    \item Adenosine substituted with adenine as the hydrolase reaction is believed to be much faster than the body response
    \item Concentration of adenine and NO kept constant for dynamic analysis due to intra- and extracellular abundance 
    \item The initial conditions used for time domain analysis correspond to maximal insufficiency based on \cite{nocon}, \cite{IL10con} and \cite{Adenosineconc}.
    \item Stochastic response ignored due to the large number of E. Coli used.
\end{itemize}
  A list of all parameters and concentrations used in figures, charts and graphs can be found at the end of the section. 
  \subsection{Negative pathway dynamics}
\begin{figure}[h]
    \centering
    \includegraphics[width = 8.5cm,height = 4cm]{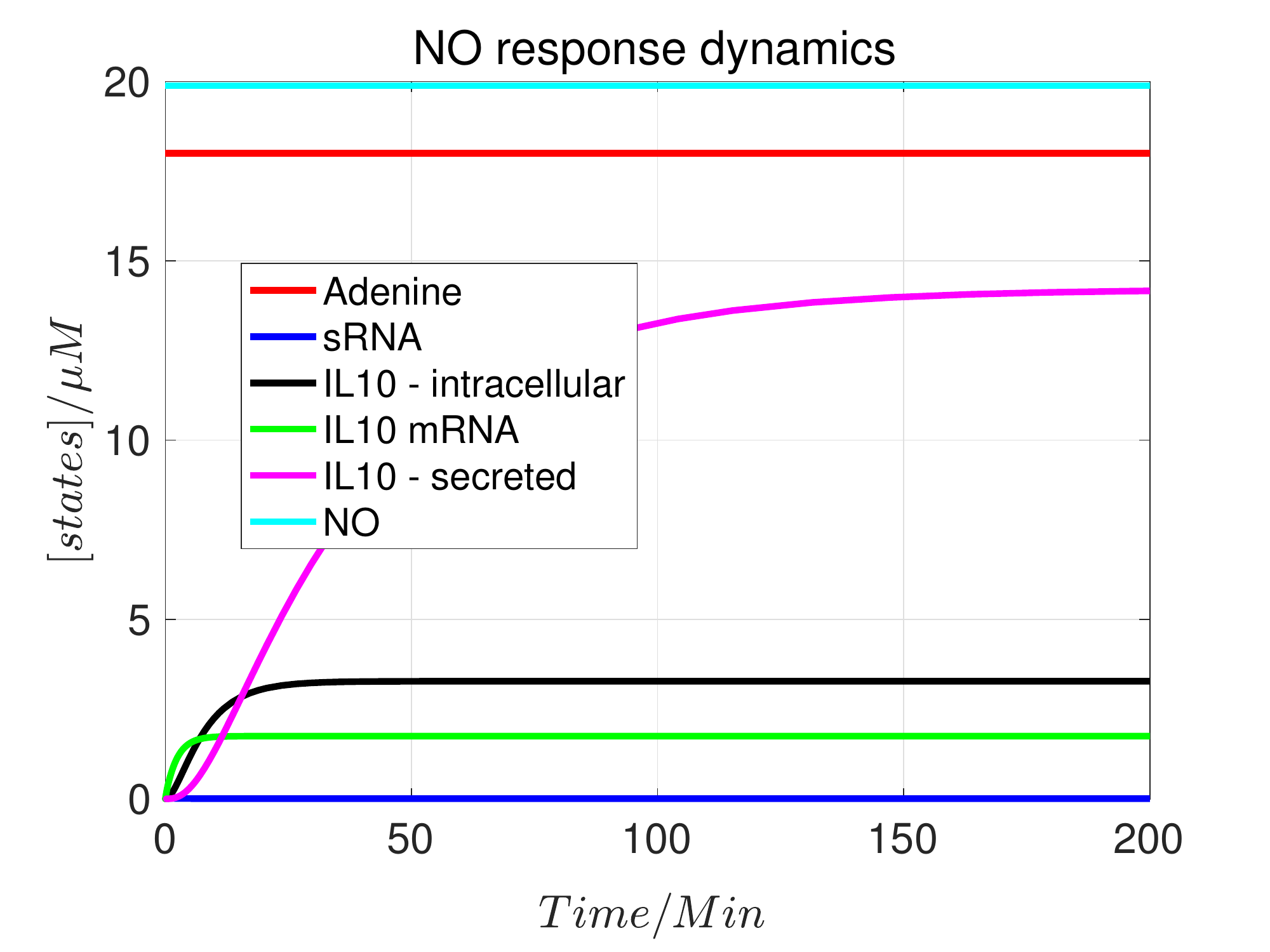}
    \caption{NO response dynamics with nominal level of Adenine}
    \label{negativepathwaydynamics}
\end{figure}

Negative pathway dynamics are fully described by the ODEs stated previously. It should be noted that based on the second assumption, the level of NO would not change with time and hence $\frac{d[NO]}{dt}= 0$. An initial concentration of $19.88 \mu M$ was used for elevated level of NO corresponding to IL-10 deficiency in patients with IBD~\cite{nocon} as well as the nominal concentration of $18 \mu M$ for adenine~\cite{Adenosineconc}. The evolution of the system to NO stimuli is illustrated in Figure~\ref{negativepathwaydynamics}.

As it can be seen from Figure~\ref{negativepathwaydynamics}, $[\text{IL-10}]$  reaches a concentration of $14.16 \; \mu M$ after roughly 180 seconds. It is important to note that the initial concentration of IL-10 is set to 0 for simplicity, meaning that in reality, the model can correct deficiencies of IL-10 to a margin of $14.16 \mu M$. A control mechanism is essential in order for the probiotic bacteria to be able to maintain a healthy level of IL-10. To determine how system would respond and whether or not it would be able to stabilise the IL-10 concentration, it is necessary to construct a large-scale version of the model that incorporates the body as well as the system embedded in a negative feedback loop~\cite{feedback}. The control modelling is discussed in more detail later in the paper. Lastly, the model behaviour is as expected due to the constant production rate, as opposed to the degradation rate, and hence a bounded steady state value is expected.
\subsection{Positive pathway dynamics}
\begin{figure}[h]
    \centering
    \includegraphics[width = 8 cm,height = 4cm]{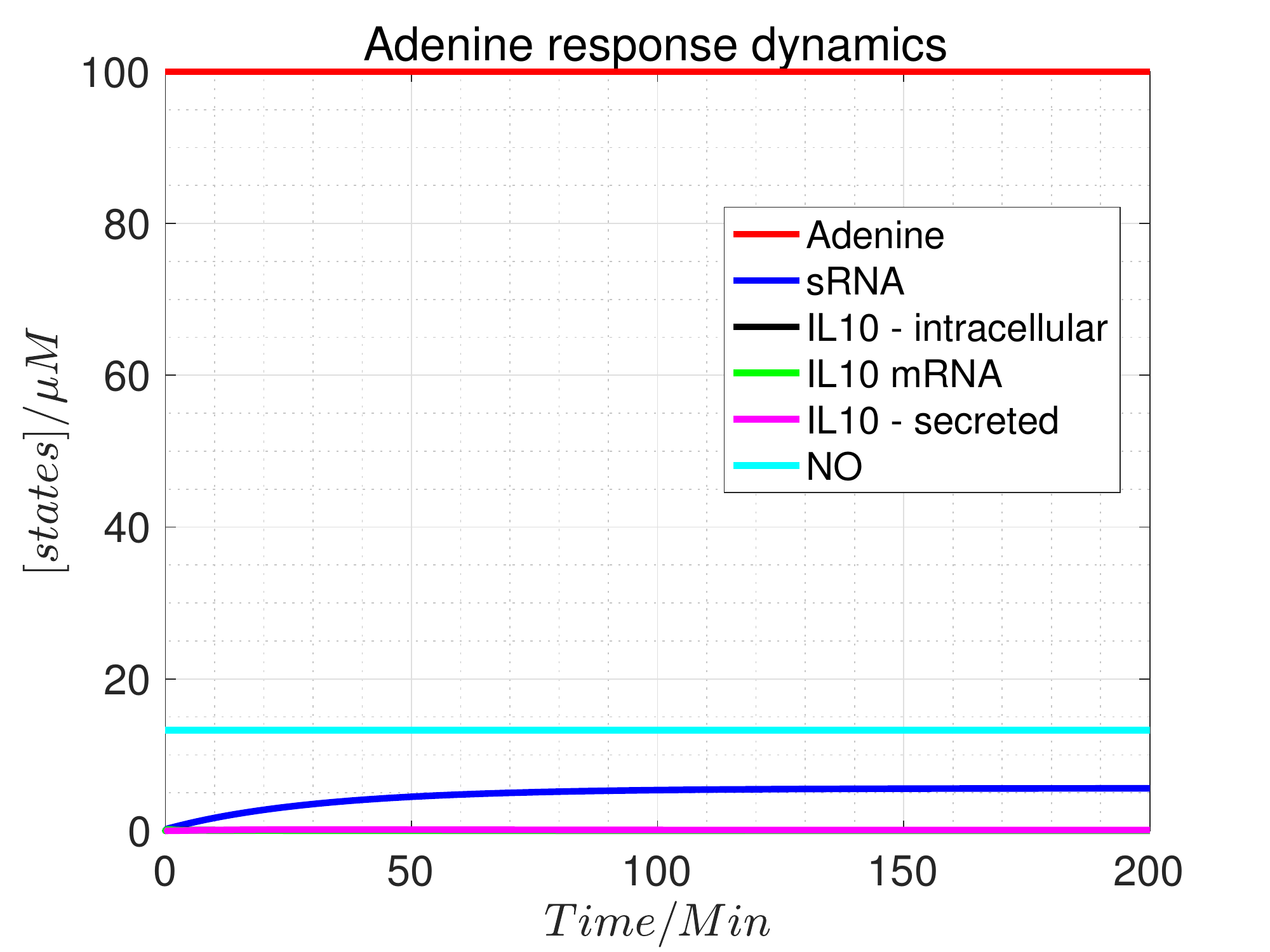}
    \caption{Adenine response dynamics with nominal level of NO}
    \label{positivepathwaydynamics}
\end{figure}

The complexity of the positive pathway is visible from Figure~\ref{schempaths}. The complexity of the system was maintained for the dynamic analysis based on ODEs stated previously. It should be noted that based on the second assumption, the concentration of adenine would not change with time and hence $\frac{d[Adenine]}{dt}= 0$. An initial concentration of $100 \mu M$ was used for elevated level of adenine corresponding to IL-10 abundance in patients with IBD~\cite{Adenosineconc} as well as the nominal concentration of $13.24 \mu M$ for NO~\cite{nocon}. The evolution of the system to adenine stimuli is illustrated in Figure~\ref{positivepathwaydynamics}.
\newline
It can be observed from Figure \ref{positivepathwaydynamics}, that $[\text{IL-10}]$ is inhibited significantly and it could be assumed that a high concentration of adenine could inhibit translation of IL-10 completely. It should be noted that the length of sRNA and the promoter strength in both negative and positive pathways($k_1\, ,\, \beta_1 \, ,\, \beta_2$) play a significant role and hence they have been optimised for maximum inhibition as well as maximum translation of IL-10 in both positive and negative pathways respectively. The sensitivity analysis and sRNA models are discussed below.
\section{Sensitivity analysis and sRNA modelling}
\begin{figure}[h]
    \centering
    \includegraphics[width = 8.5cm,height = 7cm]{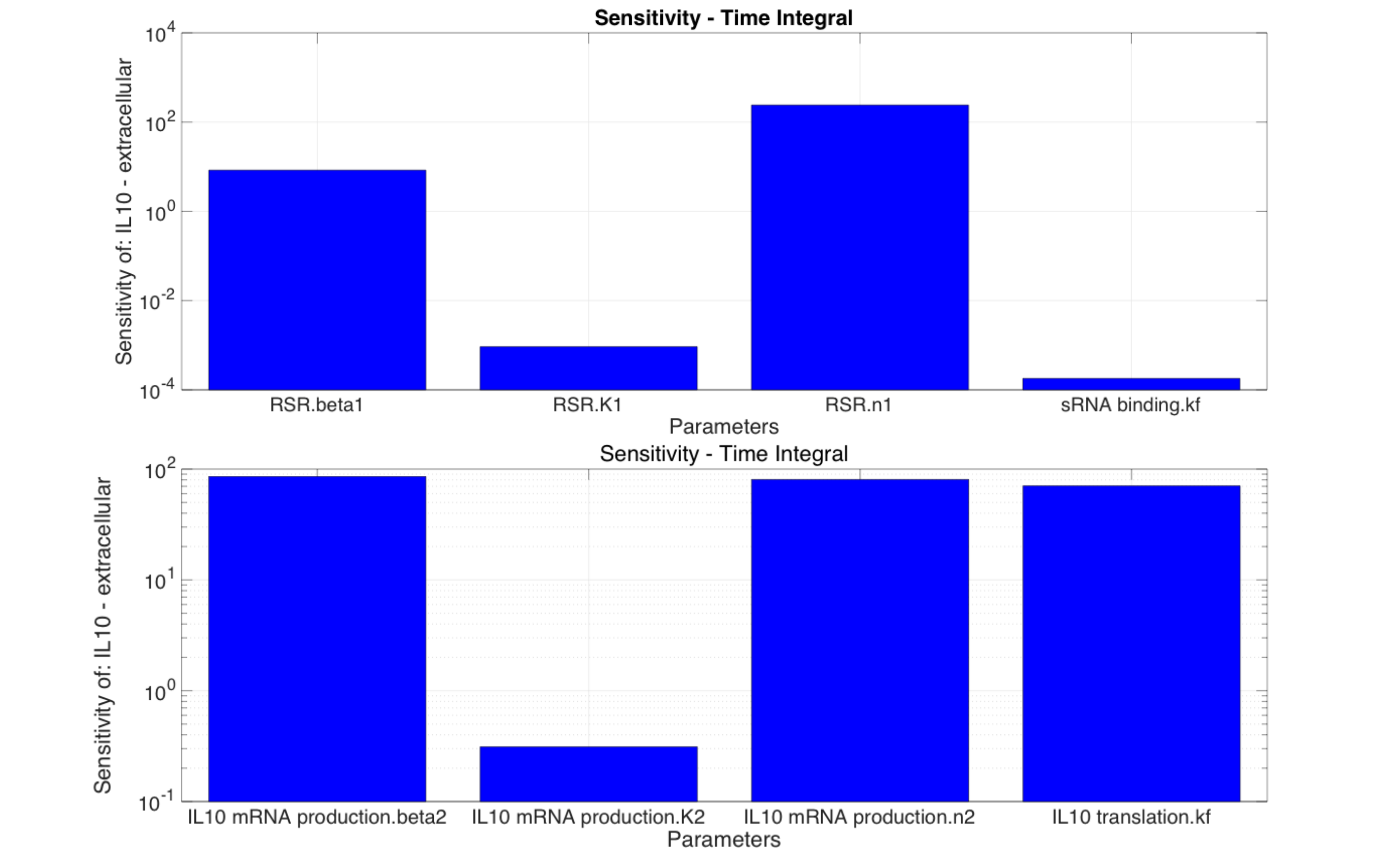}
    \caption{Sensitivity Analysis of [IL-10] with respect to parameters}
    \label{sensitivity}
\end{figure}

\begin{figure}[h]
    \centering
    \includegraphics[width = 8.5cm,height = 4.5cm]{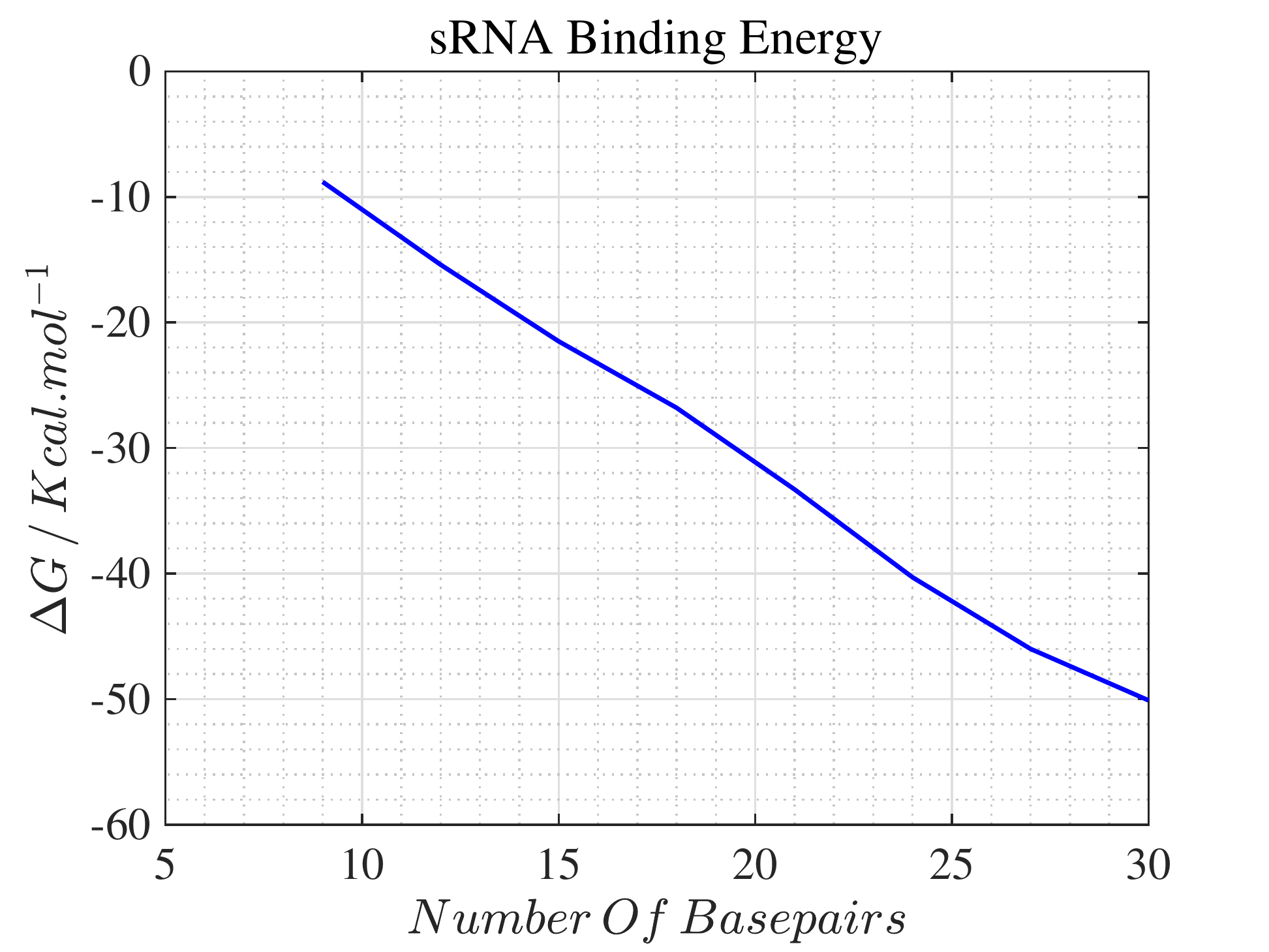}
    \caption{sRNA binding free energy variation with its base pair position}
    \label{srnaopt}
\end{figure}

Sensitivity analysis was done separately for each reaction pathway in order to find the most significant steps and parameters, thus enabling these parameters to be exploited in order to improve the responsiveness of the system to smaller perturbations in concentrations. The sensitivity of $[IL-10]$ with respect to parameters used in simulations is shown in Figure ~\ref{sensitivity}.
\newline
It is clear from Figure~\ref{sensitivity} that IL-10 production depends heavily on $\beta_2$ and $\beta_1$ which is desirable as different promoter strengths (weak, medium and strong) would enable optimisation of the response. It can also be observed that sRNA binding rate $k_f$ has less significant effect on the output compared to $\beta_2$ and hence, a stronger promoter would mean that a much longer sRNA would be required to inhibit IL-10 translation. The length of sRNA was optimised to 24 base pairs~\cite{srnabp} yielding $\Delta G = -45.97 Kcal\cdot mole^{-1}$. The graph of $\Delta G $ against the position of base pairs is illustrated in Figure~\ref{srnaopt}. 
\newline
Based on the calculated $\Delta G$,  $\beta_1$, $\beta_2$ and $k_1$ were set to $1 nMmin^{-1}$, $10 nMmin^{-1}$ and $100 nM^{-1}min^{-1}$ for complete inhibition of IL-10 translation in the positive pathway, in addition to a high IL-10 translation rate in negative pathway.
\section{Steady state model}
\begin{figure}[h]
    \centering
    \includegraphics[width = 8.5cm,height = 12cm]{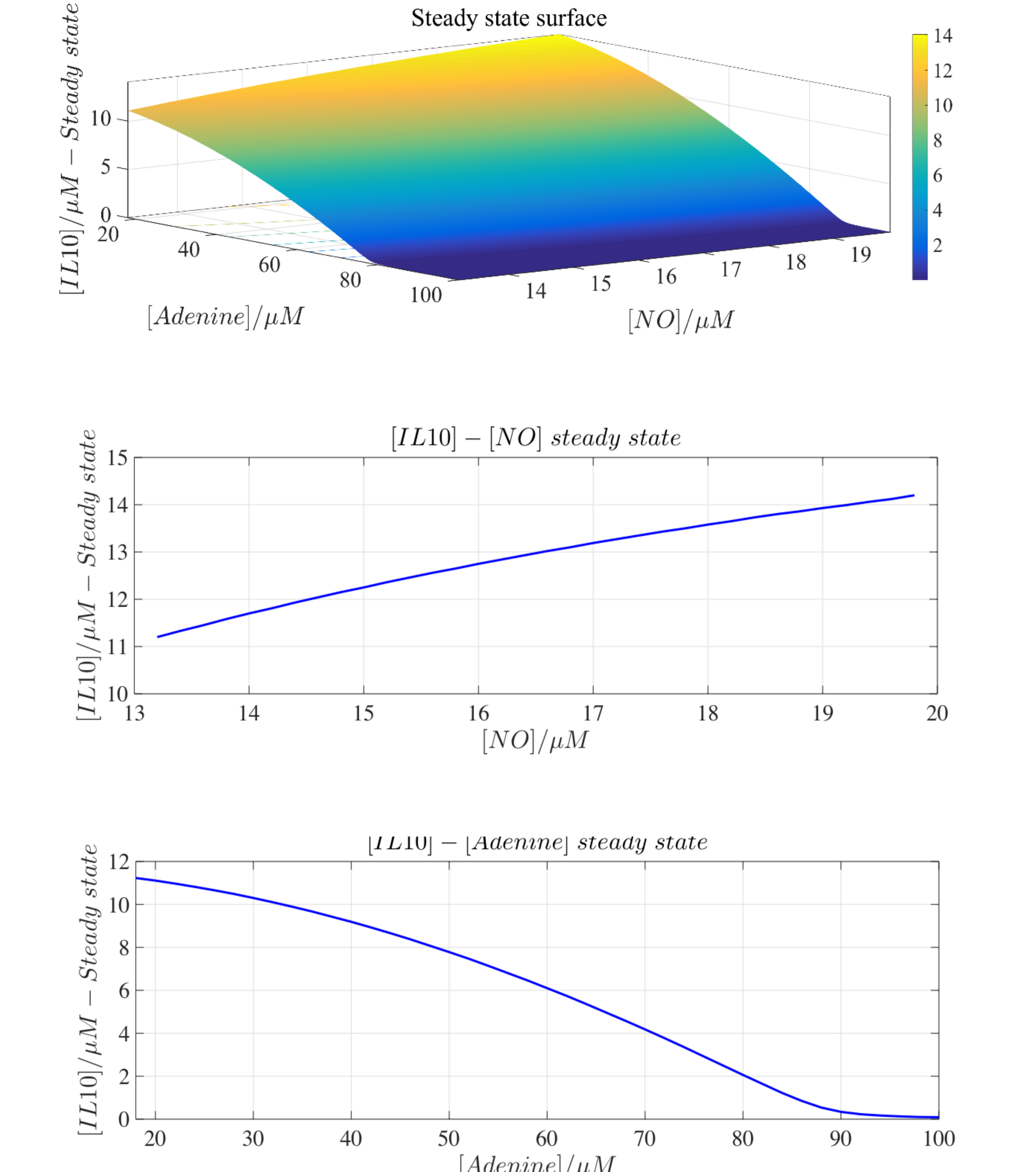}
    \caption{The steady state surface and its contours to both of the inputs}
    \label{steadystate}
\end{figure}

The simplest model is an output/input, time-independent plot which describes the system fate in response to varying input concentrations. Secretion of IL-10 has been ignored for steady state response as it is assumed that the system is given ample time to settle, secreting all translated IL-10. The steady state concentrations of species can be determined by equating (\ref{eq1}), (\ref{mRNA}) and (\ref{eq3}) to zero. The steady state values for [IL-10], [sRNA] and [IL-10 mRNA] are given below:
\begin{equation}
\label{mrnas}
    [\text{IL-10 mRNA}]^* = \frac{\alpha_2 \alpha_1 +k_1(\Gamma_1 -\Gamma_2) \pm \sqrt{\omega}}{-2\alpha_2k_1},
\end{equation}
\begin{equation}
    [\text{sRNA}]^* = \frac{\Gamma_1}{\alpha_1 +k_1[\text{IL10 mRNA}]^*},
\end{equation}

\begin{equation}
    \label{il-10s}
    [\text{IL-10}]^* = \frac{k_2}{\alpha_4}[\text{IL-10 mRNA}]^*,
\end{equation}
\newline
where $\omega = \big(k_1(\Gamma_2-\Gamma_1)-\alpha_1\alpha_2\big)^2+4\alpha_1\alpha_2 k_1 \Gamma_2$. It should be noted that $\Gamma_i$ denotes Hill functions for adenine and NO. A surface can be plotted describing the output/input steady state relation, with contours representing the steady state characteristic of each of the pathways. The plots can then be used for experimental fitting to obtain more realistic parameters. The steady state surface and it's contours could be found in Figure \ref{steadystate}.
\section{Parameters and modelling predictions for circuit design}
The parameters used for modelling are all listed below: 
\begin{itemize}
    \item Transcription level : $\beta_1=10 \; nM/min$ , $\beta_2=1 \; nM/min$ both optimised to give satisfactory results~\cite{oxfordeng}. $K_1 = 300 \; nM$~\cite{riboswitch}, $K_2 = 10 nM$~\cite{oxfordeng}.
    \item Translation level : $k_1 = 100\; min^{-1}$ optimised based on the number of base pairs in sRNA. $k_2 = 0.3 \;min^{-1}$ is the translation rate of IL-10~\cite{oxfordeng}. 
    \item Degradation and Diffusion : $\alpha_1 = 0.03 \; min^{-1}$, $\alpha_2 = 0.14 \; min^{-1}$, $\alpha_3 = \alpha_4 = 0.03 \; min^{-1}$ all based on~\cite{oxfordeng}. The diffusion rate parameter was calculated based on the number of transport proteins on the membrane of E. Coli to be $k_3 = 0.13 min^{-1}$ \cite{diffusion}.
    \item Body response : The value of $b_1$ and $b_2$ were calculated from Figure~\ref{Linearbodyresponse} and a chosen value of $a = 10^{-4}$ s. $b_1 = -1.6690$ and $b_2 = 28.24$.
\end{itemize}
\subsection{Parameters range and output dependency}
Three of the parameters used throughout the paper can be tuned for optimisation, namely: $\beta_1$, $\beta_2$ and $k_1$; some of which have been previously mentioned. These parameters were modified slightly about their optimised value to find a range for which the system maintains a desired working state. The optimised range for each of the parameters is given below. 
\begin{enumerate}
\item $\beta_1$ has a wide range of $8 < \beta_1 <90$ for which a satisfactory results could be obtained. 
\item As opposed to $\beta_1$, $\beta_2$ has a very narrow range of $0.7<\beta_2<1.3$ for which a satisfactory results could be obtained. 
\item $k_1$ is almost saturated at the value chosen for simulation and hence does not have an upper bound; however, there is a lower bound of $1<k_1$ for which a satisfactory result can be obtained.
\end{enumerate}
Equations \ref{mrnas} - \ref{il-10s} can be differentiated to find the dependency of IL-10 with respect to all three tunable parameters. As $[\text{IL10}]^*$ depends linearly on $[\text{IL10 mRNA}]^*$, (\ref{mrnas}) has been differentiated with respect to $\Gamma_1$, $\Gamma_2$ and $k_1$ for simplicity. It should be noted that since $\beta_i$ is the only tunable parameter in Hill functions, $\Gamma_i$ has been used as the parameter for simplicity in differentiation.
\begin{equation*}
    \frac{d[\text{IL-10 mRNA}]^*}{d\Gamma_1} = \frac{d}{d\Gamma_1}\big(\frac{\alpha_2 \alpha_1 +k_1(\Gamma_1 -\Gamma_2) + \sqrt{\omega}}{-2\alpha_2k_1}\big)
\end{equation*}
\begin{equation}
\label{eq18}
    \therefore \quad = \frac{k_1}{2\alpha_2k_1}\bigg(\frac{1}{\sqrt{\omega}}\big(k_1(\Gamma_2 - \Gamma_1)-\alpha_1\alpha_2\big)-1\bigg),
\end{equation}
\begin{equation*}
     \frac{d[\text{IL-10 mRNA}]^*}{d\Gamma_2} = \frac{d}{d\Gamma_2}\big(\frac{\alpha_2 \alpha_1 +k_1(\Gamma_1 -\Gamma_2) + \sqrt{\omega}}{-2\alpha_2k_1}\big)
\end{equation*}
\begin{equation}
\label{eq19}
    \therefore \quad = \frac{k_1}{2\alpha_2k_1}\bigg(1-\frac{1}{\sqrt{\omega}}\big(k_1(\Gamma_2 - \Gamma_1)-\alpha_1\alpha_2\big)\bigg),
\end{equation}
\begin{equation*}
     \frac{d[\text{IL10 mRNA}]^*}{dk_1} = \frac{d}{dk_1}\big(\frac{\alpha_2 \alpha_1 +k_1(\Gamma_1 -\Gamma_2) + \sqrt{\omega}}{-2\alpha_2k_1}\big)
\end{equation*}

\begin{equation}
\label{eq20}
    \therefore \quad = \frac{\bigg((\Gamma_1 - \Gamma_2)+ \lambda \bigg)(-2\alpha_2k_1)-2\alpha_2\mu}{(-2\alpha_2k_1)^2},
    \end{equation}

where $\lambda = \frac{\big(k_1(\Gamma_1-\Gamma_2)-\alpha_1\alpha_2\big)(\Gamma_1 - \Gamma_2\big)+4\alpha_1\alpha_2\Gamma_2}{\sqrt{\omega}} $ and $\mu =\big(\alpha_1\alpha_2+k_1(\Gamma_1 - \Gamma_2)+\sqrt{\omega}\big) $. Based on (\ref{eq18}), (\ref{eq19}) and (\ref{eq20}), the dependency of IL-10 with respect to $\beta_1$, $\beta_2$ and $k_1$ is clear. The expression $k_1(\Gamma_2-\Gamma_1)-\alpha_1\alpha_2$ is less than $\sqrt{\omega}$ and therefore it could be concluded that:
\begin{enumerate}
    \item $\frac{d[\text{IL-10 mRNA}]^*}{d\Gamma_1} <0 $
\item $\frac{d[\text{IL-10 mRNA}]^*}{d\Gamma_2} >0$
\item $\frac{d[\text{IL-10 mRNA}]^*}{dk_1} <0$
\end{enumerate}
$\beta_1$ and $k_1$ are both in the inhibitory pathway as opposed to $\beta_2$ and hence, all three statements above validate the model. 
\subsection{Modelling predictions for circuit design}
Based on the range and dependency of $[\text{IL-10}]$ on different parameters, a medium strength promoter is suggested to be used for negative pathway as opposed to positive pathway where either a strong promoter or a medium strength promoter with a high copy number is suggested. As for the length of sRNA, it was shown that as long as it has just a few base pairs, corresponding to $k_1 =1$, the output would not be affected significantly and hence, the most energetically feasible length is used.

\section{Body Dynamics}
\begin{figure}[h]
    \centering
    \includegraphics[width = 8.5cm,height = 4cm]{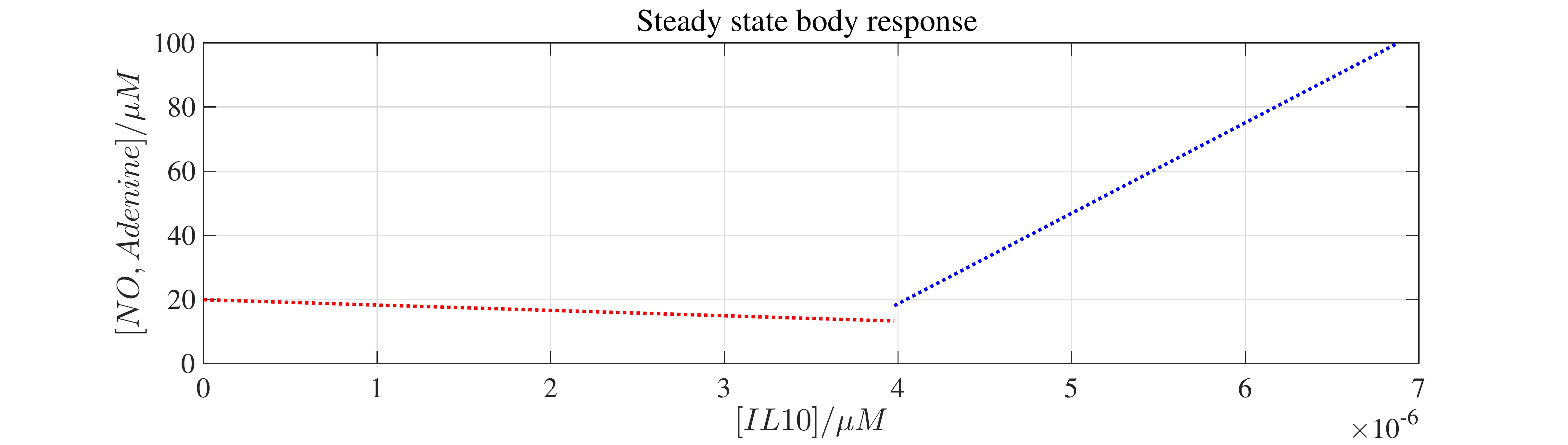}
    \caption{Linear approximation for steady-state body response in both reaction pathways}
    \label{Linearbodyresponse}
\end{figure}

\begin{figure}[h]
    \centering
    \includegraphics[width = 8 cm,height = 4.5cm]{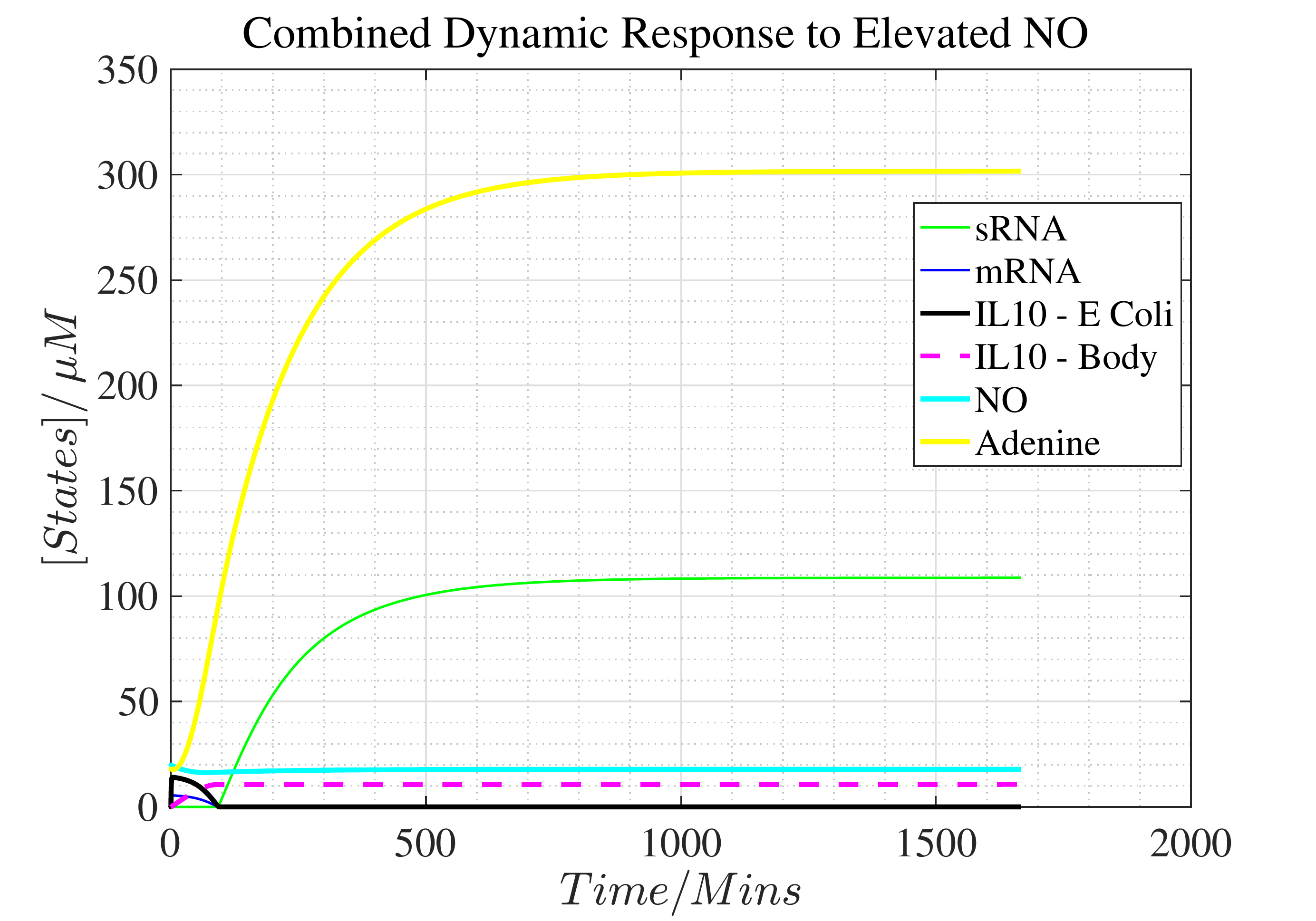}
    \caption{E. Coli - Body dynamic response to elevated level of NO}
    \label{bodydynamicsNO}
\end{figure}

\begin{figure}[h]
    \centering
    \includegraphics[width = 8 cm,height = 4.5cm]{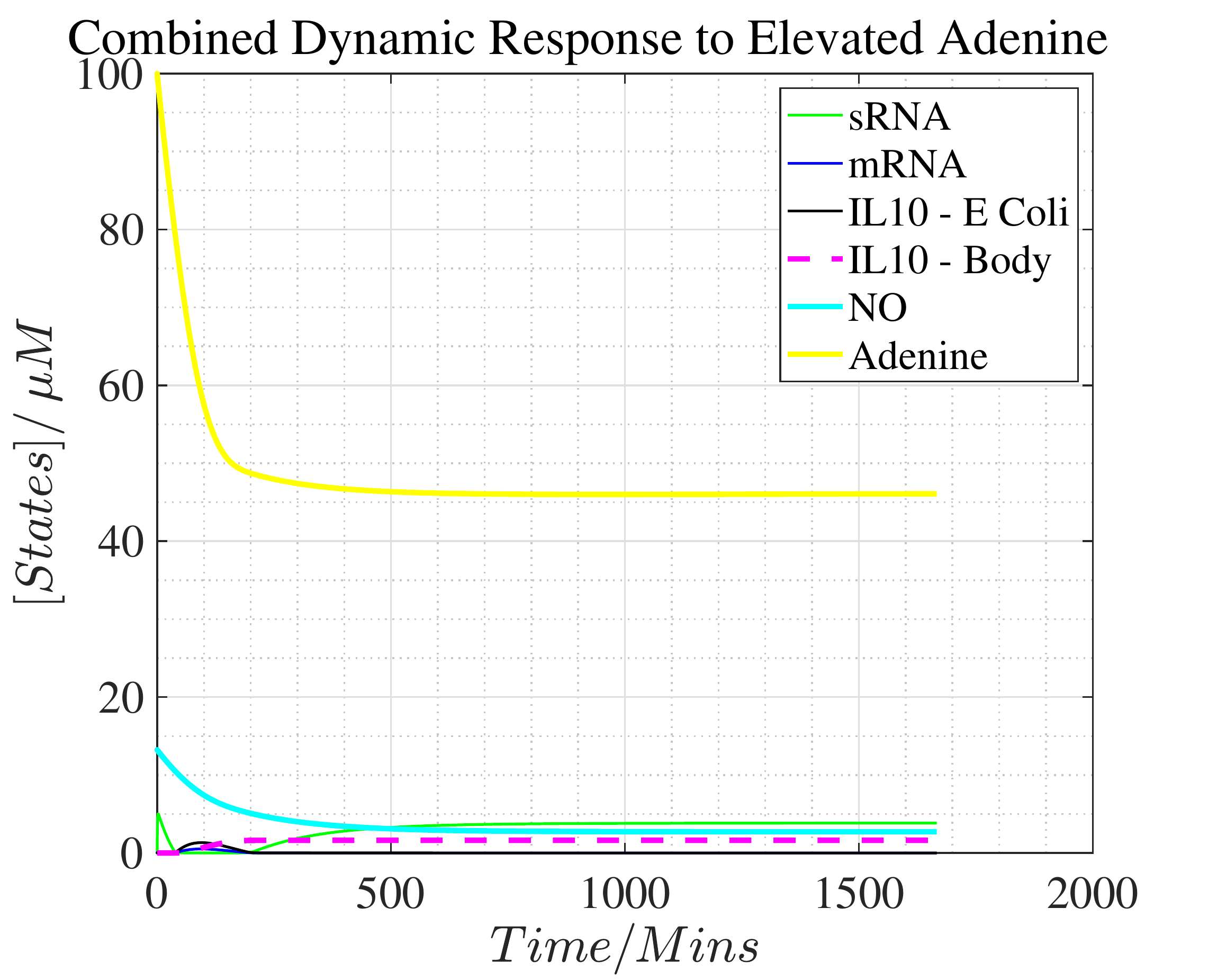}
    \caption{E. Coli - Body dynamic response to elevated level of adenine}
    \label{bodydynamicsadenine}
\end{figure}

\begin{figure}[h]
    \centering
    \includegraphics[width = 8 cm , height = 4 cm]{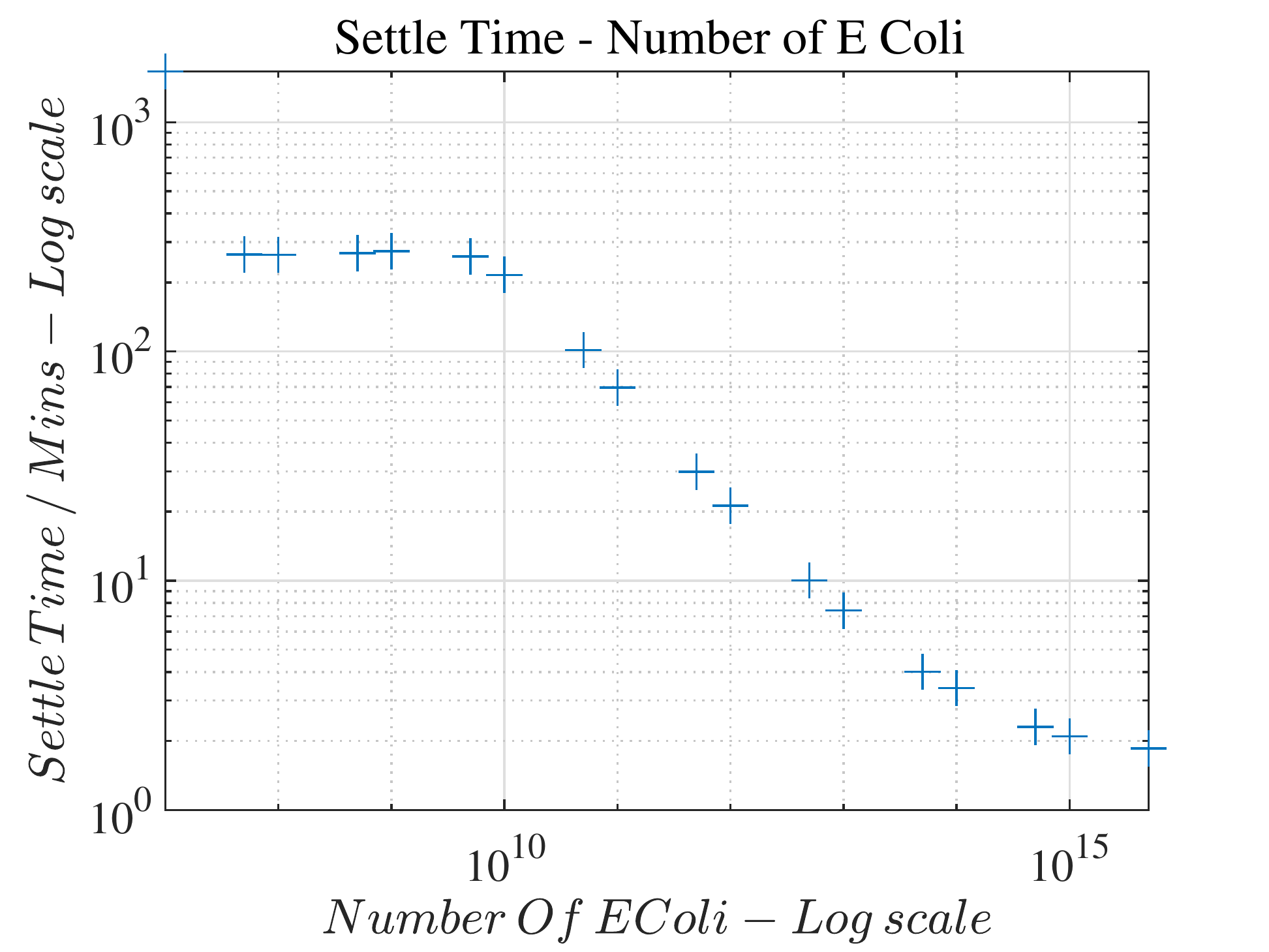}
    \caption{90 \% Settle Time - Number of E Coli used for Simulation}
    \label{numberofecoli}
\end{figure}

The response of the body is crucial to obtain comprehensive model reflecting systemic interactions between the engineered E. coli and the body. Body response mechanisms have been studied in detail however due to stochastic noise, significant assumptions were made to simplify the response for modelling. These assumptions are summarised below; please note that even though some of the assumptions are considerable, they are negligible compared to the amount of variation present in bodily systems and between individuals:
\begin{itemize}
    \item The model is based on bulk amount of bacteria translating or inhibiting IL-10.
    \item The body responds in about 3-4 hours and the response has been taken to be linear due to lack of data points.
    \item The body has been assumed to behave similarly to an LTI (Linear time-invariant) system and the model can be extrapolated, with care, for people with different body responses.
\end{itemize}
Based on assumptions stated above, two lines can be plotted for elevated levels of NO and Adenosine versus IL-10 - the highest and lowest concentration of IL-10 ~\cite{IL10concentration} and its corresponding signalling molecule~\cite{nocon}~\cite{Adenosineconc} that has been measured in human body. The plot of steady-state concentrations is shown in Figure~\ref{Linearbodyresponse}.
\newline
The dynamic response of the body is assumed to be linear with a response time of about 3 hours$(10^{4} s)$ as very few data points were available for the steady state response of the body. Therefore, the differential equation describing the response is as follows:
\begin{equation}
    \frac{d[\text{NO,Adenine}]}{dt} = -a_i[\text{NO,Adenine}]+b_i[\text{IL10}]+c_i,
    \label{bodyode}
\end{equation}
Figure~\ref{Linearbodyresponse} can be used to find $\frac{b}{a}$ and $\frac{c}{a}$ for both positive and negative pathways. Then by choosing $ a = 10^{-4} s^{-1}$, for the response time, (\ref{bodyode}) is fully defined and can be solved simultaneously with (\ref{eq1}) - (\ref{eq4}) to give an overall prediction on how the body and the probiotic bacteria would interact with each other to reach an equilibrium concentration of IL-10 in the body. It should be noted that for these two to be solved together, one should be aware of the correction factor required to approximate metabolite concentrations inside a bacterium relative to the body, in addition to an indication of the number of bacteria being used. Hence, the initial number of bacteria is optimised in a correction loop by looking at the settle time of IL-10 inside the body. The dynamics of IL-10 in the body for elevated level of NO and adenine are shown in Figures \ref{bodydynamicsNO} and \ref{bodydynamicsadenine} respectively. It should be noted that for representation purposes, the concentration of IL-10 in the body - magenta dashed line - has been multiplied by $10^{6}$.
\newline
It is clear from Figures above that the body and E. Coli reach a stable concentration of IL-10 without oscillation in any of the metabolite concentrations. The accuracy of the response time of $10^{4}$ seconds is reflected by the similarity of the steady-state. 
\newline
The number of E Coli used for simulation would significantly affect the settle time or the response time and hence, the time taken by the combined system to reach $90 \%$ of the steady-state concentration of IL-10 has been plotted against the number of bacteria in Figure \ref{numberofecoli}. It is essential to understand the significance and importance of the positive pathway in stability and robustness of the proposed probiotic. Hence, dynamics of the combined system has been plotted in absence of Adenine pathway to demonstrate that the system becomes unstable without a negative feedback loop.  
\begin{figure}[h]
    \centering
    \includegraphics[width = 8 cm , height = 4.5 cm]{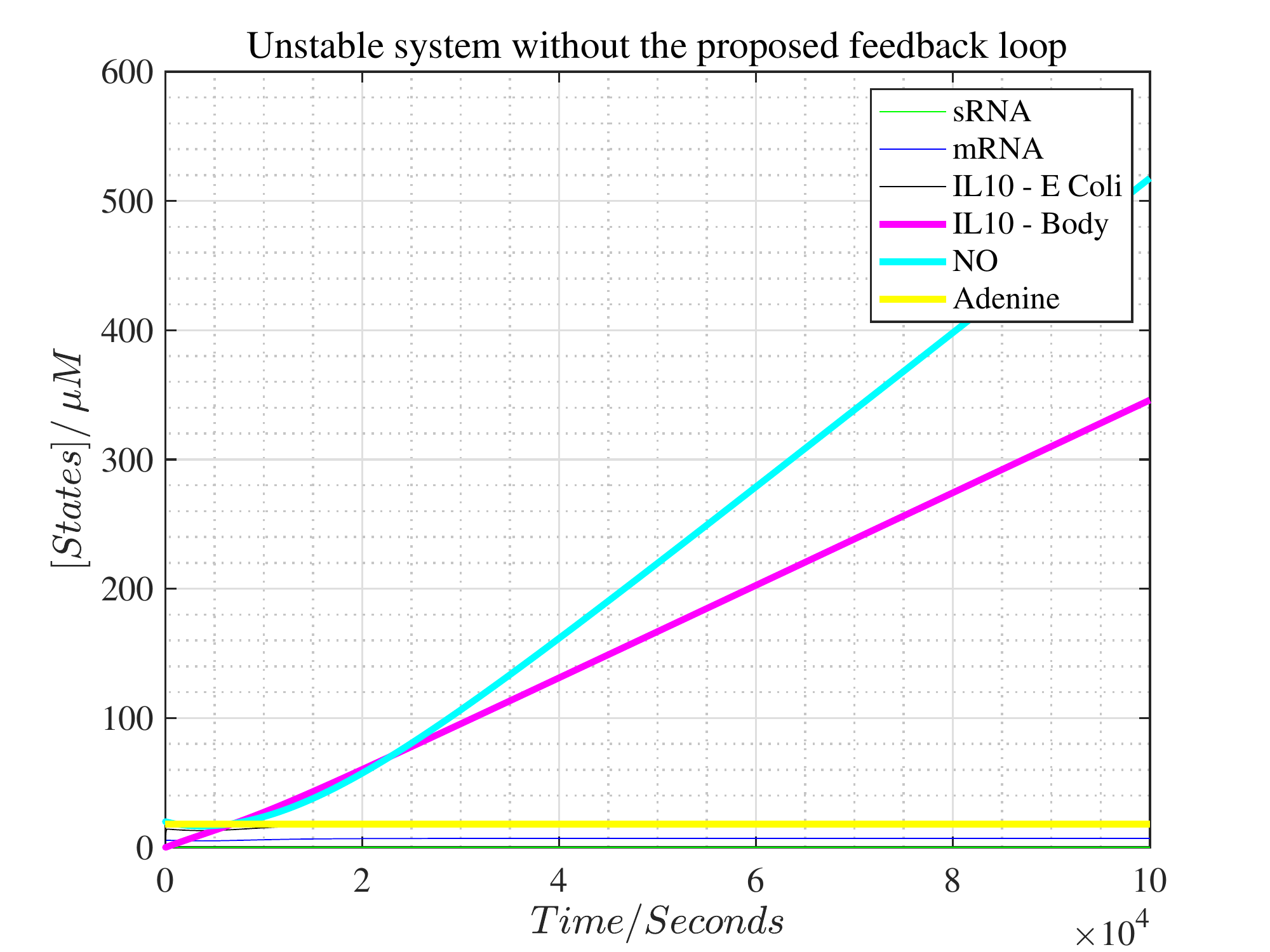}
    \caption{Unstable system in the absence of the inhibiting pathway}
    \label{unstablesystem}
\end{figure}
\newpage
\section{Negative feedback and frequency domain analysis}
\begin{figure}[h]
    \centering
    \includegraphics[width = 8 cm]{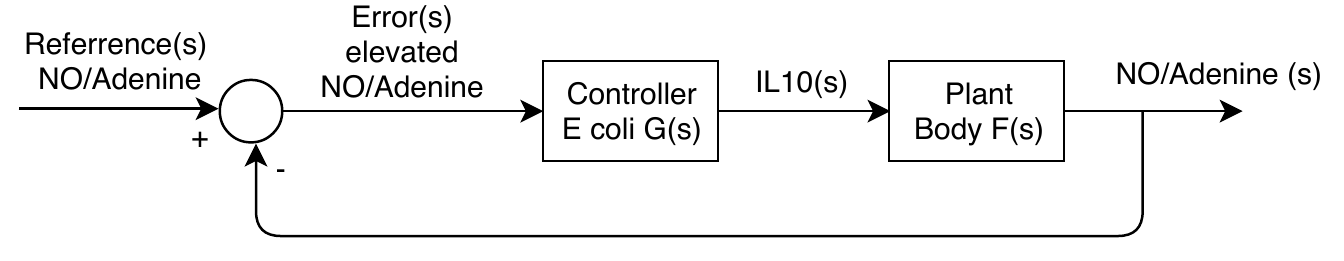}
    \caption{Cascade compensation design}
    \label{Cascadecompensation}
\end{figure}

As discussed earlier in the paper, both positive and negative feedback could potentially correct the concentration of IL-10 in the body. However, it is important to note that body response should be included in the model to give an actual and meaningful prediction of how E. coli would behave in the body. For this purpose, control theory and the negative feedback loop have been used by transforming the body and the bacteria in a standard cascade compensation model and deriving their transfer functions. Hence, we developed a cascade compensation model where we have the body as the plant and the bacteria as the controller. Assumptions made for frequency domain analysis are listed below:
\begin{itemize}

    \item The system has been linearised about an equilibrium point and therefore, it has been assumed that perturbations are small for the Taylor series to converge. 
    \item The reaction pathways will be analysed separately as based on superposition, the response of the linearised model would be a sum of its response to inputs separately. 
    \item Secretion has been ignored for simplicity in transfer function. 
    
\end{itemize}
Therefore, by using the correct transfer function, both models would fall into the same control design, just with different transfer functions. The proposed design for the feedback loop could be seen in Figure~\ref{Cascadecompensation}.
\newline
The aim of the controller is to set the reference signal, namely NO and adenosine, to nominal level as this would correspond to a healthy concentration of IL-10 within the body. Transfer functions for the controller- both positive and negative pathway- and the body are required for control analysis of system response in a human body.

\section{Transfer functions}
\begin{figure}[h]
    \centering
    \includegraphics[width = 8.5cm,height = 4cm]{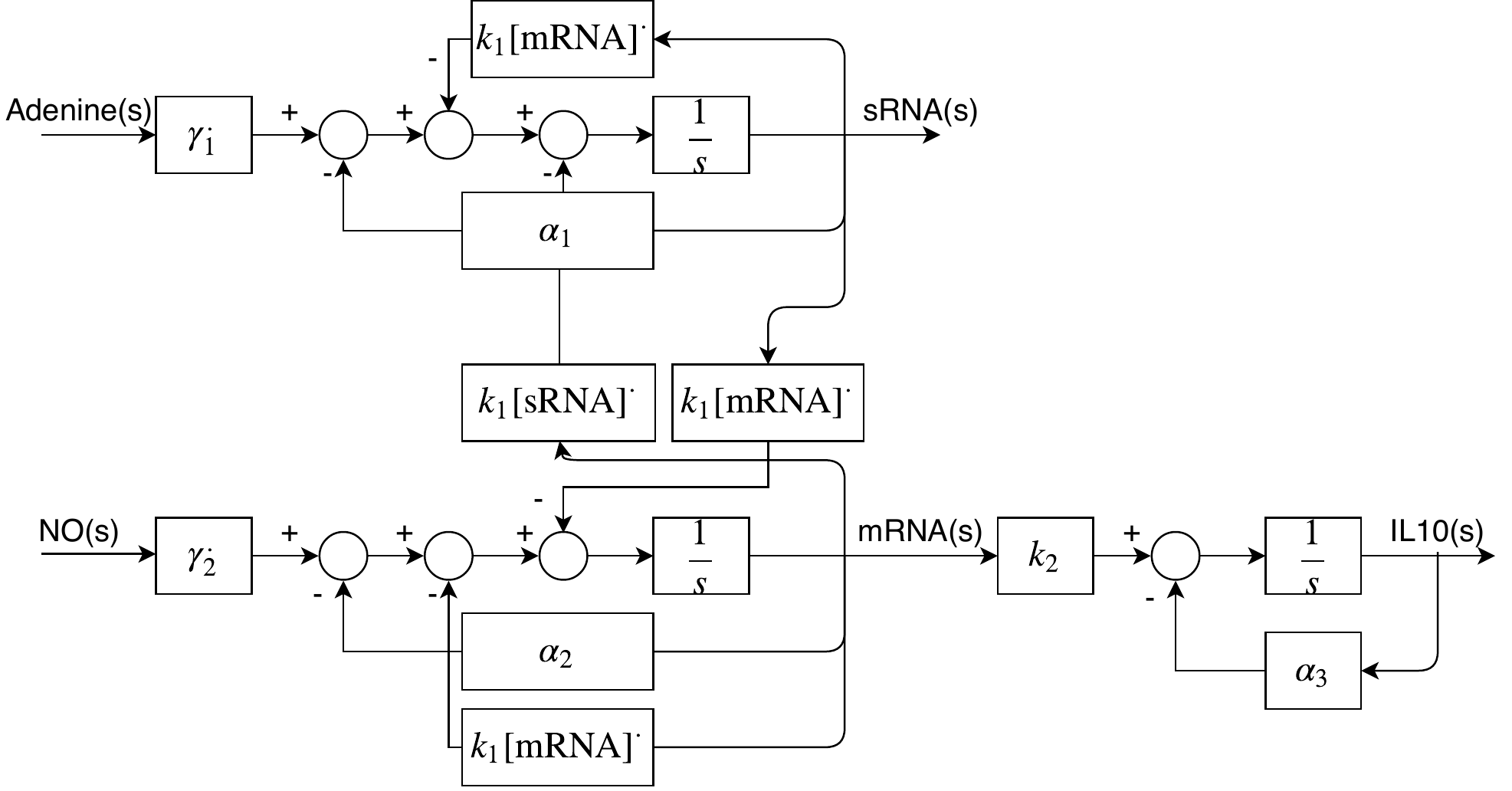}
    \caption{Block diagram in frequency domain for linearised model}
    \label{block}
\end{figure}

\begin{figure}[h]
    \centering
    \includegraphics[width = 8.5cm,height = 5cm]{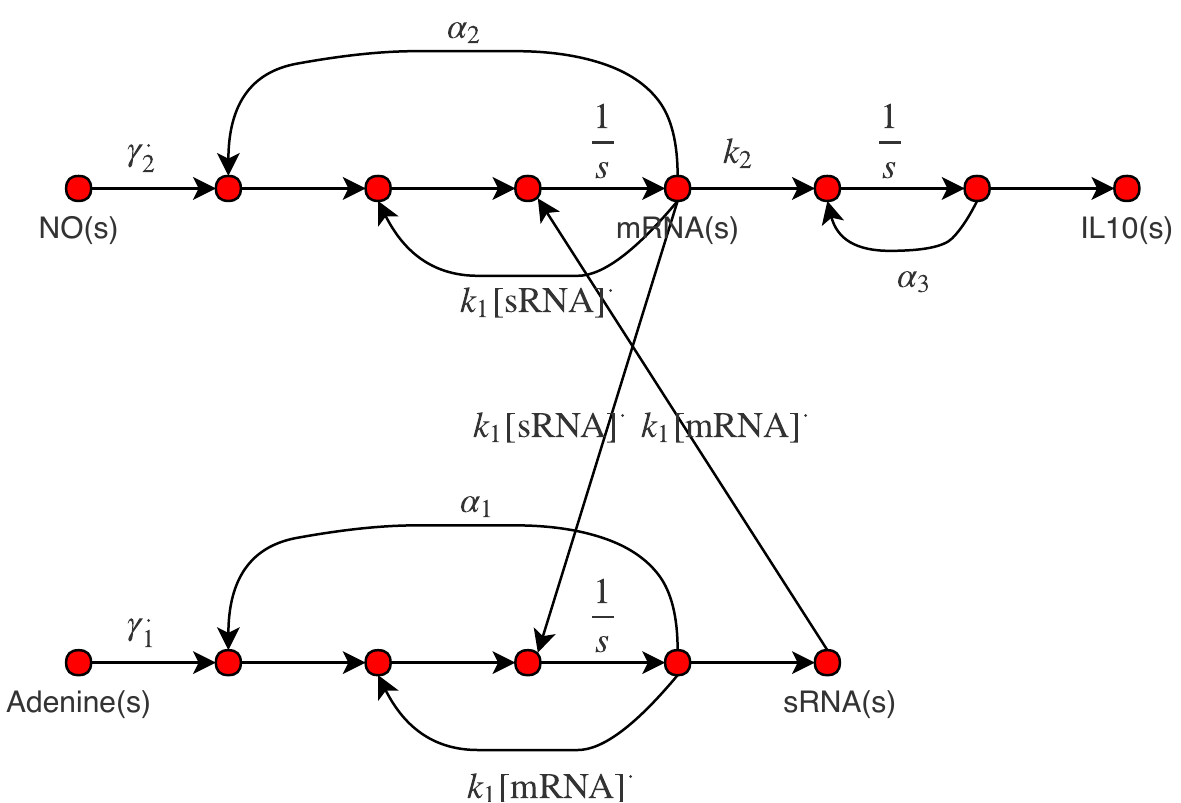}
    \caption{Signal Flow Graph(SFG) for the linearised model}
    \label{sfg}
\end{figure}

As stated in the assumptions, the model has been linearised due to non-linearity raised by the Hill functions and sRNA binding. Therefore, Hill functions have been replaced by $\gamma_i^*$ as stated earlier in Equation \ref{linearhill}. The block diagram and signal flow graph for the linearised system are shown in Figures \ref{block} and \ref{sfg}, where transcription of IL-10 mRNA, sRNA and translation of IL-10 can be seen clearly.
\newline
Transfer functions for both negative and positive pathways were found using Mason's gain relation~\cite{mason} and confirmed against~\cite{oxfordeng}. Transfer functions of both positive and negative pathways can be found below:
\begin{equation}
    G_1(s) = \frac{\text{IL10(s)}}{\text{Adenine(s)}} = \frac{k_2\gamma_1^*(s+\alpha_1+k_1[\text{IL10 mRNA}]^*)}{(s+\alpha_3)(s+s_+)(s+s_-)},
\end{equation}
\begin{equation}
    G_2(s) = \frac{\text{IL10(s)}}{NO(s)} = \frac{-k_1[\text{IL10 mRNA}]^*k_2\gamma_2^*}{(s+\alpha_3)(s+s_+)(s+s_-)},
\end{equation}
\newline
where $s_\pm = \frac{1}{2}(\alpha_2 + k_1[\text{sRNA}]^* + \alpha_1 +k_1[\text{IL10 mRNA}]^*)\pm \frac{1}{2}\sqrt{d}$ and $d = (\alpha_2 + k_1[\text{sRNA}]^*- \alpha_1-k_1[\text{IL10 mRNA}]^*)^2+4k_1^2[\text{IL10 mRNA}]^*[\text{sRNA}]^*$. 
\subsection{Body response transfer function}
Laplace transform can be taken from both sides of the Equation \ref{bodyode} and by ignoring $c$, being very small and negligible, the transfer function for the body can be derived:
\begin{equation*}
s[\text{NO,Adenine(s)}] = -a_i[\text{NO,Adenine(s)}]+b_i[\text{IL10(s)}]
\end{equation*}
\begin{equation}
    \implies F_i(s) =\frac{[\text{NO,Adenine(s)}]}{[\text{IL10(s)}]} = \frac{b_i}{s+a_i}
\end{equation}
\section{Stability}
\begin{figure}[h]
    \centering
    \includegraphics[width = 8.5cm,height = 4cm]{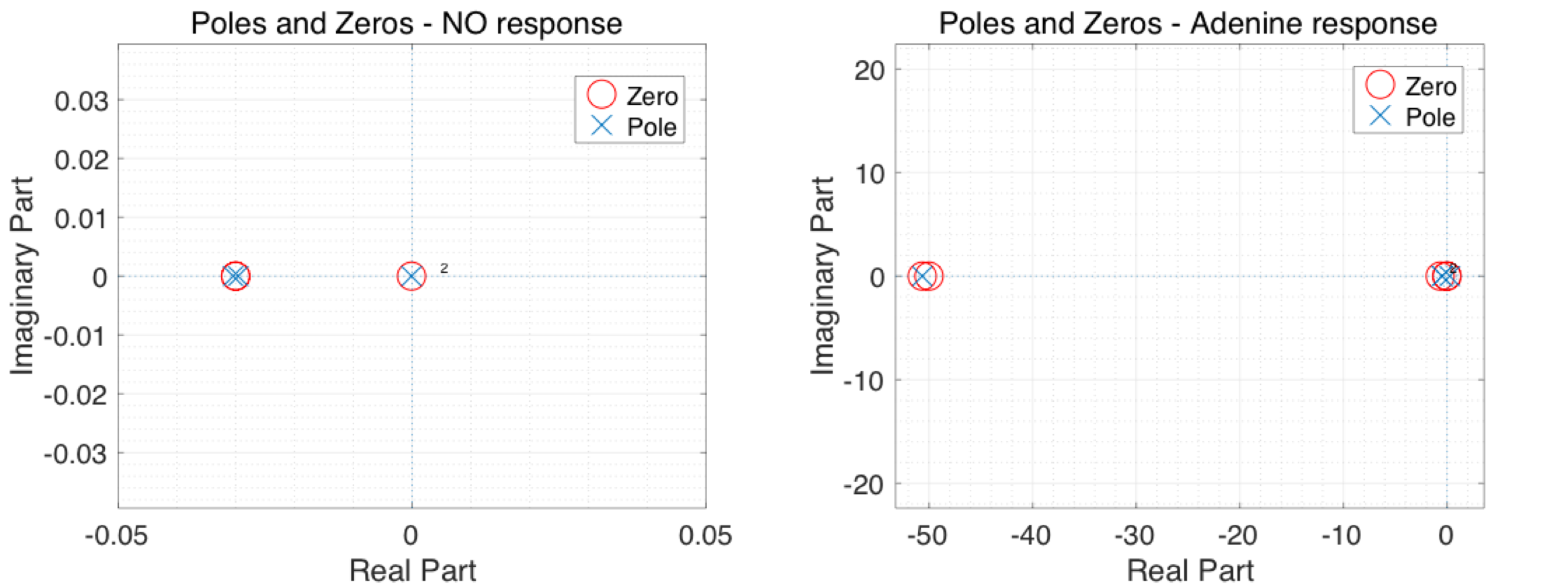}
    \caption{poles and zeros for NO and Adenine responses }
    \label{argand}
\end{figure}

The closed loop transfer function for cascade cascade controller design given in Figure~\ref{Cascadecompensation} would contain important information regarding steady state error and relative stability. It should be mentioned that similar to most control problems where the controller is designed for specific specifications~\cite{control}, the reaction pathway can be changed by altering promoter strength and sRNA length, making it almost possible to adopt desired structures for the controller. The closed loop transfer function for positive and negative pathways is as follows:
\begin{equation}
\label{closeloop}
    T_i(s) = \frac{F_i(s)G_i(s)}{1+F_i(s)G_i(s)}
\end{equation}
\newline
Based on (\ref{closeloop}), the stability of the system can be determined by observing the location of the poles which are roots of the characteristic polynomial $1+F_i(s)G_i(s) = 0$. The characteristic equation for adenine and NO reference signals is given in Equations \ref{char1} and \ref{char2}.
\begin{equation}
\label{char1}
    1+\frac{k_2\gamma_1^*(s+\alpha_1+k_1[\text{IL10 mRNA}]^*)}{(s+\alpha_3)(s+s_+)(s+s_-)}\cdot \frac{b_1}{s+a_1} =0
\end{equation}
\begin{equation}
\label{char2}
    1+\frac{-k_1[\text{IL10 mRNA}]^*k_2\gamma_2^*}{(s+\alpha_3)(s+s_+)(s+s_-)} \cdot \frac{b_2}{s+a_2} =0
\end{equation}
\newline
Equations (\ref{char1}) and (\ref{char2}) were solved numerically~\cite{poles} and smallest poles and zeros for both reaction pathways are summarised in the Argand diagram below, Figure \ref{argand}. All poles and Zeros are summarised after the figure.  
\newline
It should be noted that some poles and zeros were not plotted as they were significantly larger(in magnitude) than those shown in Figure \ref{argand} and would have thus hindered comparison with those illustrated. All poles and zeros are given below. 
\begin{enumerate}
    \item NO response : 
    \begin{itemize}
        \item Poles : $s_1 \approx -1.075\times10^4$ , $s_2 \approx -0.030$ , $s_3 \approx -0.030$ and $s_4 \approx −1.060\times10^{−4}$.
        \item Zeros : $z_1 = 1.075\times10^4$, $z_2 = -0.030$, $z_3 = -0.030$, $z_4 = 1.000\times10^{-4}$.
    \end{itemize}
    \item Adenine response :
    \begin{itemize}
        \item Poles : $s_1 \approx −50.656$, $s_2 \approx −0.474$, $s_3 \approx −0.474$ and $s_4 =  \approx 0.262$.
        \item Zeros : $z_1 = −50.656$, $z_2 = −50.000$, $z_3 = −0.656$, $z_4 = −1.000\times 10^{−4}$.
    \end{itemize}
\end{enumerate}
Despite right-plane poles, it should be noted that both responses can be considered stable as the dominant poles are significantly larger- in magnitude- than all other poles. 
\section{Steady state error - DC gain}
\begin{equation*}
    E(s) = \frac{1}{1+G_i(s)F_i(s)}R(s) \; , \; R(s) = \frac{C}{s} \quad C \in \mathbb{R}
\end{equation*}
\begin{equation*}
   FVT \, :\quad \lim_{t\rightarrow \infty} e(t) = \lim_{s \rightarrow 0} sE(s)
\end{equation*}
\begin{equation}
    \therefore \; \lim_{t \rightarrow \infty} e(t) = \lim_{s\rightarrow 0} \frac{1}{1+G_i(s)F_i(s)} 
    \end{equation}
By substituting back Transfer functions for both E. Coli,$G_i(s)$, and the body,$F_i(s)$, the steady state error can be found as follows: 
\begin{itemize}
    \item For NO response $\lim_{t\rightarrow\infty} e(t) = -2.9501\times 10^{-4} $.
    \item For adenine response $\lim_{t\rightarrow\infty} e(t) = -2.1912 \times 10^{-3} $.
\end{itemize}
It is evident that the combined model has a very small DC gain and hence, a very small margin of error is estimated for the steady concentration of IL-10.

\section*{Acknowledgement}
Supervision and guidance of the following people were significant in achieving and finishing this paper. Their contributions and help is greatly appreciated.
\begin{itemize}
\item Karandip Saini - Department of Chemistry, our team wet Lab coordinator
    \item Dr. Aivar Sootla - Department of Engineering sciences University of Oxford.
    \item Dr.Nicolas Delalez - Department of Engineering sciences University of Oxford.
    \item Prof.Antonis Papachristodoulou - Department of Engineering sciences University of Oxford. 
    \item Prof.Christopher Macminn - Department of Engineering sciences University of Oxford.
    \item Dr.George Wadhams - Department of Biochemistry University of Oxford.
\end{itemize}

\newpage
\medskip
\bibliographystyle{unsrt}
\bibliography{Arman}

\end{document}